\documentclass[12pt,a4paper]{article}

\usepackage{a4wide}
\usepackage{amsmath,amssymb,amsfonts,amsxtra,simpler-wick,mathrsfs} 
\usepackage{hyperref}
\usepackage{physics}
\usepackage{caption}
\usepackage{subcaption}
\usepackage[nosort,noadjust]{cite}
\usepackage{cleveref}
\allowdisplaybreaks

\begin{document}
\thispagestyle{empty}
\vspace*{-2.5cm}
\begin{center}
\hfill BONN--TH--2024--16
\vskip 0.6in
{\Large
\bf Half-Wormholes in a Supersymmetric SYK Model}
\end{center}

\vspace*{1cm}

\centerline{Stefan F{\"o}rste\footnote{forste@th.physik.uni-bonn.de}
and Saurabh Natu\footnote{snatu@uni-bonn.de}}
\vspace{1cm}

\begin{center}{\it
Bethe Center for Theoretical Physics\\
{\footnotesize and}\\
Physikalisches Institut der Universit\"at Bonn,\\
Nussallee 12, 53115 Bonn, Germany}
\end{center}

\centerline{\bf Abstract}
\vskip .3cm

We identify half-wormhole contributions to the non averaged 
$\mathscr{N} =1$ supersymmetric SYK model in which time has been reduced to a point.
As in previously studied examples, the inclusion of half-wormholes restores factorisation in the large $N$ limit. Wormholes as well as half-wormholes break supersymmetry.

\newpage
\section{Introduction}
\label{sec:intro}

In the context of AdS/CFT duality \cite{Maldacena:1997re} the factorisation problem is related to wormhole geometries contributing to the gravitational partition function \cite{Maldacena:2004rf}. The partition function depending on boundary values of the fields does not factorise into disconnected component contributions due to wormholes connecting them. In the dual description each connected boundary component corresponds to a separate CFT. These CFT's are decoupled from each other. The lack of factorisation can be explained by considering a statistical ensemble of CFT's instead \cite{Coleman,Giddings_Strominger,Marolf_Maxfield_Model}. Lower dimensional examples are pure JT gravity \cite{JT_theory_Teitelboim,JT_theory_Jackiw} being dual to a random matrix ensemble  \cite{Saad_JT_matrix} and its supersymmetric extensions \cite{Stanford:2019vob,Turiaci:2023jfa}.  There are three dimensional examples as well (see e.g.\ \cite{Afkhami-Jeddi:2020ezh,Maloney:2020nni} and references thereof).
JT gravity is also dual to the low energy limit of the SYK model 
which is an ensemble too since couplings are random \cite{Backreaction,Sachdev_Ye,Kitaev_lec,Kitaev_Soft_mode,Maldacena:2016hyu,Jensen_Chaos_SYK,Engelsoy_backreaction,Cvetic:2016eiv}. 
In the large $N$ limit, the saddle point approximation becomes exact in the SYK model.
One can associate a given saddle point to a wormhole geometry. If one does not average over couplings factorisation is restored on the SYK side. 
Saddle points whose contributions are lost when averaging can restore factorisation upon their inclusion. Technically, the authors of \cite{saad2021wormholes} identified a nice way of obtaining such saddle points. To simplify the problem the time interval had been reduced to an instant.  
Then, the fourth moment of the partition function was computed. The authors of \cite{saad2021wormholes} determined saddle points which do not contribute to the second moment squared. They called these saddle points half-wormholes. In a non averaged theory computing the $n$th moment and taking the $n$th power is the same. Including half-wormholes into the computation of the second moment squared should therefore be a step towards restoring factorisation. Indeed, it turns out that the inclusion of half-wormholes suffices. 
This has been confirmed by a direct computation of the fixed coupling expression \cite{Mukhametzhanov:2021nea}.
More work on extensions and applications to modified models can be found in 
\cite{Mukhametzhanov:2021nea, Choudhury:2021nal, Blommaert:2021gha,
  Garcia-Garcia:2021squ, Saad:2021uzi, Goto:2021mbt,
  Mukhametzhanov:2021hdi, Blommaert:2021fob, Goto:2021wfs,
  Das:2022uhj, Peng:2022pfa, Cheng:2022nra, Tian:2022zpc}.
  
In the present paper, we will consider a supersymmetric version of the SYK model \cite{SUSY_SYK}. It has been argued in \cite{N1_AdS2_Iris, N2_AdS2} that for the low energy limit the bulk dual is given by supersymmetric JT gravity \cite{Chamseddine:1991fg}. Gravitational wormholes contribute to the bulk partition function \cite{Stanford:2019vob,Turiaci:2023jfa}.
We will see, that in a simplified version of the ${\cal N}=1$ supersymmetric SYK model the inclusion of half-wormholes restores factorisation. The paper is organised as follows. In \Cref{sec: Avg SYK model} we introduce the simplified version of the model and compute the second moment of the partition function. After a contour deformation the same result can be reproduced in a saddle point approximation. If it were not for the wormholes the second moment should vanish. Therefore, the saddle point configuration  can be viewed as a holographic dual of a wormhole, or for short, as a wormhole. In \Cref{sec: Fixed calculations} half-wormhole contributions are determined by studying the fourth moment of the partition function. It is argued that their inclusion restores factorisation. In \Cref{SUSY_breaking} we look at supersymmetry and find it to be broken by wormholes as well as half-wormholes. \Cref{sec: Conclusion} states our conclusions. Some computational details are shifted into appendices.  

\section{Averaged \texorpdfstring{$\mathbf{{\cal N}=1}$}{TEXT} SYK model}\label{sec: Avg SYK model}
 
 The supersymmetric SYK model can be defined in terms of $N$ fermions $\psi_i$ ($i \in \left\{ 1, \ldots, N\right\}$).
 Its supercharge is given by \cite{SUSY_SYK}
 \begin{equation}
 Q =\,\,\,\,\,\,\text{i}\!\!\!\!\!\!\!\!\!
 \sum_{1\leq i<j<k<l\leq N} \!\!\!C_{ijk}\psi^i\psi^j\psi^k,
\end{equation}
where the totally anti-symmetric  couplings $C_{ijk}$ are drawn from a Gaussian ensemble with
\begin{equation}
\left< C_{ijk}\right> = 0 \,\,\, ,\,\,\, \left< C_{ijk}^2\right> = \frac{2J}{N} ,
\end{equation}
and $J$ is a dimensionfull (energy) constant which is kept finite in the large $N$ limit. This model is supersymmetric because the Hamiltonian is given by the supercharge squared. 
Following the authors of \cite{saad2021wormholes} we will mostly consider a restricted setting in which the fields $\psi_i$ are time independent. The time interval is replaced by an instant at $t=1$.  Then Hamiltonian and Lagrangian differ only by a sign. To simplify the computation of averages it is useful to linearise the Hamiltonian in the random couplings by introducing auxiliary bosonic fields $b_i$ ($i \in \left\{ 1,\ldots ,N\right\}$) \cite{SUSY_SYK}. The Lagrangian is given by
\begin{equation}
L = -\frac{1}{2}b^i b^i  + \text{i}\sum_{i<j<k}C_{ijk}\left(b^i \psi^j\psi^k \right).
\label{eq:lagrangian}
\end{equation}
where a sum over repeated indices is implied.
The auxiliary field can be eliminated by solving its equation of motion
\begin{equation}
b^i = \text{i}\sum_{i<j<k}C_{ijk}\psi^j\psi^k  .
\end{equation}
The Lagrangian (\ref{eq:lagrangian}) is invariant under supersymmetry transformations
\begin{equation}
\delta\psi_i =\epsilon b_i \,\,\, ,\,\,\, \delta b_i = 0 .
\label{eq:elesusy}
\end{equation}
\subsection{Evaluation of \texorpdfstring{$\mathbf{\left<Z^2\right>}$}{TEXT}}\label{subsec:saddlept calc Z^2}
As in the non supersymmetric case the ensemble average of the partition function vanishes. The average for the square of the partition function yields a finite result showing  a lack of factorisation due to taking the ensemble average.
For fixed couplings the partition function squared is given by
\begin{equation}
Z^2 = \frac{1}{\left( 2\pi\right)^N}\int d^{2N} b \, d^{2N} \psi \,\text{e}^{- L_{LR}} ,
\end{equation}
with
\begin{equation}
    L_{LR} = -\frac{1}{2}\left(b^i_Lb^i_L+b^i_Rb^i_R\right) + \text{i}\sum_{i<j<k}C_{ijk}\left(b^i_L\psi^j_L\psi^k_L+b^i_R\psi^j_R\psi^k_R\right) .
\end{equation}
Here, the subscripts $L$ and $R$ label two copies of the Lagrangian (\ref{eq:lagrangian}) 
and the path integral is performed over both sets of fields. The subscripts can be thought of as belonging to the left and right boundary of $AdS_2$.
The ensemble average is
\begin{equation}\label{eq:prime_Z^2eq}
       \left<Z^2\right> = \frac{1}{\left( 2\pi\right)^N}\int d^{2N} b d^{2N}\psi \,
       \text{e}^{-\overline{L}_{LR}}
\end{equation}
where
\begin{equation}
\overline{L}_{LR}=-\frac{1}{2}\left(b^i_Lb^i_L+b^i_Rb^i_R\right) + \frac{NJ}{2}\!\left(\frac{b^i_Lb^i_R}{N}\right)\!
      \left(\frac{\psi^j_L\psi^j_R}{N}\right)^2 \!\! - 2NJ\!\left(\frac{b^i_L\psi^i_R}{N}\right)\!
     \left(\frac{\psi^j_Lb^j_R}{N}\right)\!
     \left(\frac{\psi^k_L\psi^k_R}{N}\right) .
\end{equation}
We introduce composite quantities
\begin{equation}
G^{LR}_{\psi\psi} = \frac{\psi^j_L\psi^j_R}{N}\,\,\, ,\,\,\, G^{LR}_{bb} = \frac{b_L^j b_R^j}{N}\,\,\, ,\,\,\, 
G^{LR}_{b\psi} = \frac{b_L^j\psi_R^j}{N}\,\,\,
,\,\,\, G^{LR}_{\psi b} = \frac{\psi_L^j b^j_R}{N} .
\label{eq:composits}
\end{equation}
To the integrand in (\ref{eq:prime_Z^2eq}),
one adds an integration over these composites and an insertion 
of a  delta function restricting them to their 
values (\ref{eq:composits}). These delta functions are represented by their Fourier integrals
\begin{equation}
\delta\left( G - M/N\right) = \frac{N}{4 \pi \text{i}}\int _{\text{i}{\mathbb R}}d\Sigma\, \text{e}^{-\Sigma\left( NG -M\right)/2}
\label{eq:deltaf}
\end{equation}
for bosonic fields $G^{LR}_{\psi\psi}$, $\Sigma^{LR}_{\psi\psi}$, 
$G^{LR}_{bb}$, $\Sigma^{LR}_{bb}$. Whereas one has
\begin{equation}
\delta\left( G - M/N\right) = \frac{2}{N}\int d\Sigma\, \text{e}^{-\Sigma\left( NG -M\right)/2}
\end{equation}
for anti-commuting fields $G^{LR}_{b\psi}$, $\Sigma^{LR}_{b\psi}$, 
$G^{LR}_{\psi b}$, $\Sigma^{LR}_{\psi b}$.
We arrive at
\begin{multline}
\left< Z^2\right> = -\frac{1}{\left(2\pi\right)^{N+2}}\int d^{2N}b d^{2N}\psi \int_{\text{i}{\mathbb R}} d^4 \Sigma \int_{\mathbb R} d^4 G \,
\text{exp}\left\{ \frac
{1}{2}\left(b^i_Lb^i_L+b^i_Rb^i_R\right) + \frac{NJ}{2}\left(G^{LR}_{bb}
\right)\left(G^{LR}_{\psi\psi}\right)^2 \right.\\ \left.
- NJ G^{LR}_{b\psi}G^{LR}_{\psi b} G^{LR}_{\psi\psi}
-\frac{N}{2}\Sigma^{LR}_{\psi\psi}
\left(G^{LR}_{\psi\psi}-\frac{\psi^i_L\psi^i_R}{N}\right)-\frac{N}{2}\Sigma^{LR
}_{bb}\left(G^{LR}_{bb}-\frac{b^i_Lb^i_R}{N}\right)
\right.\\ \left.
-\frac{N}{2}\Sigma_{b\psi}^{LR}\left( G_{b\psi}^{LR}-\frac{b^i _L\psi^i _R}{N}\right)
-\frac{N}{2}\Sigma_{\psi b}^{LR}\left( G_{\psi b}^{LR}
-\frac{\psi^i _L b^i _R}{N}\right)
\right\}.
\label{eq:both}
\end{multline}
In the following we will drop the $LR$ labels at the fields since they are redundant in the present discussion.
Integrating out the $b$'s and the $\psi$'s gives rise to the following Berezinian
\begin{equation}
s\text{det}^{-\frac{N}{2}}\left( \begin{array}{cccc}
1 & \frac{\Sigma_{bb}}{2} & 0 & \frac{\Sigma_{b\psi}}{2} \\
\frac{\Sigma_{bb}}{2} & 1 &\frac{\Sigma_{\psi b}}{2} & 0 \\
0 & -\frac{\Sigma_{\psi b}}{2}&0 & \frac{\Sigma_{\psi\psi}}{2}  \\
-\frac{\Sigma_{b\psi}}{2} & 0 & -\frac{\Sigma_{\psi\psi}}{2} & 0
\end{array}\right)   = \text{e}^{-\frac{N}{2}
\text{log} \left(1 -\frac{\Sigma_{bb}^2}{4} 
+\frac{\Sigma_{\psi b}\Sigma_{b\psi}\Sigma_{bb}}{2 \Sigma_{\psi\psi}}\right)
+ N \text{log}\frac{\Sigma_{\psi\psi}}{2}},
\end{equation}\label{eq:sdet z^2}
where  we have set squares of anti-commuting quantities to zero. 
After plugging this into (\ref{eq:both}) we get
\begin{multline}
\left< Z^2\right> = -\frac{1}{4\pi^2} \int_{\text{i}{\mathbb R}} 
d^4 \Sigma \int_{\mathbb R} d^4 G \,
\text{exp}\left\{ 
-\frac{N}{2}
\text{log} \left(1 -\frac{\Sigma_{bb}^2}{4} 
+\frac{\Sigma_{\psi b}\Sigma_{b\psi}\Sigma_{bb}}{2 \Sigma_{\psi\psi}}\right)
\right. \\ \left.
+ N \text{log}\frac{\Sigma_{\psi\psi}}{2}
 + \frac{NJ}{2}G_{bb}
\left(G_{\psi\psi}\right)^2 
- NJ G_{b\psi}G_{\psi b} G_{\psi\psi}\right. \\ \left. 
-\frac{N}{2}\Sigma_{\psi\psi}
G_{\psi\psi}-\frac{N}{2}\Sigma_{bb} G_{bb}
-\frac{N}{2}\Sigma_{b\psi} G_{b\psi}
-\frac{N}{2}\Sigma_{\psi b}G_{\psi b}
\right\}.
\label{eq:effective}
\end{multline}

The integral in (\ref{eq:effective}) can be computed by first integrating over the anti-commuting degrees of freedom, then over the commuting $\Sigma$'s and finally over the commuting $G$'s. The detailed calculation is presented in appendix \ref{sec:int} with the result (in the large $N$ limit)
\begin{equation}
\left< Z^2\right> \approx 4 \sqrt{\frac{2}{3}} 2^{-N}\, 3^{\frac{3N}{4}}\, J^{\frac{N}{2}}\, \text{e}^{-N} .
\label{eq:final}
\end{equation}
%
\subsection{Identifying Wormholes}\label{sec: Identifying Wormholes}

On the gravity side of the AdS/CFT correspondence the non vanishing 
result for the expectation value of $Z^2$ is due to wormhole configurations 
connecting the two boundaries of $AdS_2$ \cite{Saad_JT_matrix,Stanford:2019vob}. On the SYK side, we would like to identify corresponding field configurations which are responsible for the non vanishing result in (\ref{eq:final}). 
The order in which we have integrated over the fields in the previous subsection does not lead to such an interpretation. Following \cite{saad2021wormholes} we deform our integration path such that the resulting
integrations can be carried out in any order. Our starting point is (\ref{eq:effective}). 
Again we first integrate over the non commuting collective fields to obtain the two contributions in \eqref{eq: Z^2 after integrating non commmuting fields}. Then, we redefine,
\begin{equation}
\Sigma_{bb} = \text{i}\sigma_b \,\,\, ,\,\,\, 
G_{\psi\psi} =\text{e}^{\frac{\text{i}\pi}{4}} g_{\psi} \,\,\, ,\,\,\, 
\Sigma_{\psi\psi} = \text{i}\text{e}^{-\frac{\text{i}\pi}{4}} \sigma_{\psi}\,\,\,\, G_{bb} = g_b .
\label{eq:def1}
\end{equation}
The deformed contour of integration is obtained by taking $g_b$, $\sigma_b$, $g_{\psi}$,
$\sigma_{\psi}$ real. We arrive at
\begin{equation}\label{eq: Z^2 two integrals after grassmann integration}
\left<Z^2\right> = \frac{N^2}{16\pi^2}\int d^2\sigma\, d^2 g\, \text{e}^{{\cal L}_1} + \frac{N^2 J}{32\pi^2}\int d^2\sigma\, d^2 g\, \text{e}^{{\cal L}_2} 
\end{equation}
with
\begin{align}\label{eq:lagrangians 1 and 2}
{\cal L}_1 = & -\frac{N}{2}\text{log} \left( 1 + \frac{\sigma_b^2}{4}\right)
+ N\text{log}\left(\frac{\text{i}\text{e}^{-\frac{\text{i}\pi}{4}}\sigma_\psi}{2}\right)
+ \frac{\text{i}NJ}{2} g_b g_\psi^2
-\frac{\text{i}N}{2}\sigma_\psi g_\psi
-\frac{\text{i}N}{2} \sigma_b g_b,\\
{\cal L}_2 = & -\left(\frac{N}{2}+1\right)\text{log} \left( 1 + \frac{\sigma_b^2}{4}\right)
+ (N-1)\text{log}\left(\frac{\text{i}\text{e}^{-\frac{\text{i}\pi}{4}}\sigma_\psi}{2}\right) + \log i\sigma_b g_{\psi}
+ \frac{\text{i}NJ}{2} g_b g_\psi^2\nonumber\\
&-\frac{\text{i}N}{2}\sigma_\psi g_\psi -\frac{\text{i}N}{2} \sigma_b g_b .
\end{align}
Next we perform the integration first over the $g_b$ and then over the $\sigma_b$ fields to reduce \eqref{eq: Z^2 two integrals after grassmann integration} to
\begin{equation}\label{eq: reduced Z^2 two integrals after grassmann integration}
\left<Z^2\right> = \frac{N}{4\pi}\int d\sigma_{\psi}\, d g_{\psi}\, \text{e}^{{\cal L}_1} + \frac{NJ}{8 \pi}\int d\sigma_{\psi}\, d g_{\psi}\, \text{e}^{{\cal L}_2} 
\end{equation}
with
\begin{align}\label{eq: reduced lagrangians 1 and 2}
{\cal L}_1 = & -\frac{N}{2}\text{log} \left( 1 + \frac{J^2 g^4_{\psi}}{4}\right)
+ N\text{log}\left(\frac{\text{i}\text{e}^{-\frac{\text{i}\pi}{4}}\sigma_\psi}{2}\right)
-\frac{\text{i}N}{2}\sigma_\psi g_\psi,\\
{\cal L}_2 = & -\left(\frac{N}{2}+1\right)\text{log} \left( 1 + \frac{J^2 g^4_{\psi}}{4}\right)
+ (N-1)\text{log}\left(\frac{\text{i}\text{e}^{-\frac{\text{i}\pi}{4}}\sigma_\psi}{2}\right) + \log\text{i}J g^3_{\psi}-\frac{\text{i}N}{2}\sigma_\psi g_\psi .
\end{align}
The saddle point equations for $g_\psi$ from ${\cal L}_1$ result in
\begin{equation}\label{eq:gp1}
-\frac{g_{\psi}^3 J^2}{\left(\frac{g_{\psi}^4 J^2}{4}+1\right)}- i \sigma_{\psi} = 0 .
\end{equation}
From taking the $\sigma_{\psi}$ derivative we learn that
\begin{equation}
\frac{1}{\sigma_{\psi}}-\frac{i g_{\psi} }{2}=0 .
\label{eq:sp1}
\end{equation}
Solving \eqref{eq:gp1} and \eqref{eq:sp1} leads to four solutions labelled by $m\in \left\{ 0,1,2,3\right\}$
\begin{align}\label{eq: L1 saddles}
g_\psi &= \frac{\sqrt{2}}{3^{\frac{1}{4}} \sqrt{J}}\text{e}^{\text{i}\pi\left( \frac{1}{4} -\frac{m}{2}\right)},\nonumber\\
\sigma_\psi &= -3^{\frac{1}{4}} \sqrt{2J}\text{e}^{\text{i}\pi\left( \frac{1}{4} +\frac{m}{2} \right)}.
\end{align}
Similarly, the four solutions for $g_{\psi}$ and $\sigma_{\psi}$ resulting from the saddle point equations for ${\cal L}_2$ are
\begin{align}\label{eq: L2 saddles}
g_\psi &= \frac{\sqrt{2} \left(N-4\right)^{\frac{1}{4}}}{(3N)^{\frac{1}{4}} \sqrt{J}}\text{e}^{\frac{\text{i}\pi}{2}\left( 1 -m \right)},\nonumber\\
\sigma_\psi &= 
\frac{ \sqrt{2J} 3^{\frac{1}{4}}N^{\frac{1}{4}}}{\left( N - 4\right)^{\frac{1}{4}}}\left( 1 - \frac{1}{N}\right)\text{e}^{\frac{\text{i}\pi m}{2}} .
\end{align}
The only terms in the Lagrangians yielding different contributions for different solutions are  $\log \sigma_\psi $ and $\log g_\psi$ terms in ${\cal L}_1$ and  term in ${\cal L}_2$ respectively. That is, labelling the Lagrangian evaluated at a particular solution by $m$ one finds the following structure
\begin{equation}
{\cal L}^{m}_{1/2} = {\cal L}_{1/2} + 2\pi \text{i} N\left(\frac{m}{4}\right)
\end{equation}
with,
\begin{align}
    {\cal L}_1 &= -N + \frac{N}{2}\log\left(\frac{3^{\frac{3}{2}}}{2}\right) + N\log \sqrt{J}\\
    {\cal L}_2 &= -N + \frac{N}{2}\log\left(\frac{3^{\frac{3}{2}}}{2}\right) + N\log \sqrt{J} + \log \frac{2}{J}\\
\end{align}
The sum over $m \in \left\{ 0,1,2,3\right\}$ yields an overall factor
\begin{equation}
1 + \text{e}^\frac{\text{i}\pi N}{2} + \text{e}^{\text{i}\pi N}  + \text{e}^\frac{3\text{i}\pi N}{2} .
\end{equation}
For integer $N$ this is non zero only if $N$ is also divisible by four. The observation that $\left< Z^2\right>$ 
is non vanishing only if $N$ is a multiple of four agrees with our previous calculation.
Taking into account the contribution from the one-loop factor for the two integrals in \eqref{eq: reduced Z^2 two integrals after grassmann integration} (see appendix \ref{ap:loop}) we find that the contribution from one saddle point is
\begin{equation}
\left< Z^2\right> \approx \sqrt{\frac{2}{3}}2^{-N}3^{\frac{3N}{4}}J^{\frac{N}{2}} \text{e}^{-N} .
\end{equation}
Thus, we see that all four saddle points contribute such that the sum over all saddle points is equal to the previous 
evaluation (\ref{eq:final}).

\subsection{Computing \texorpdfstring{$\mathbf{\left<Z^4\right>}$}{TEXT}}\label{N1 Z4}

Half-wormholes are collective field configurations forming certain saddle points
contributing to $\left<Z^4\right>$ \cite{saad2021wormholes}. Therefore, we consider now the fourth power of the partition function
\begin{equation}
    Z^4 = Z_LZ_RZ_{L'}Z_{R'} = \frac{1}{\left( 2 \pi\right)^{2N}}\int\dd[4N]{\psi} \dd[4N]{b}e^{-L}.
\end{equation}
with,
\begin{align}
    L = \frac{1}{2}\left( \left( b^i_L\right)^2\right. & \left. +\left(b^i_R\right)^2+\left(b^i_{L'}\right)^2+\left(b^i_{R'}\right)^2\right)
    \nonumber \\
    &-\text{i}\sum_{i<j<k}C_{ijk}\left( b^i_L\psi^j_L\psi^k_L+b^i_R\psi^j_R\psi^k_R +b^i_{L'}\psi^j_{L'}\psi^k_{L'}+ b^i_{R'}\psi^j_{R'}\psi^k_{R'}\right) .
\end{align}
The average over the couplings is,
\begin{equation}
        \left<Z^4\right> = \frac{1}{\left( 2 \pi\right)^{2N}}\int \dd[4N]{\psi}\dd[4N]{b} e^{-\Bar{L} }
\end{equation}
with
\begin{equation}
    \Bar{L} = \frac{1}{2}\left(\displaystyle\sum_{\gamma}(b^i_{\gamma})^2\right) +\frac{NJ}{2}\left(\frac{b_{\alpha}^ib_{\beta}^i}{N}\right)\left(\frac{\psi_{\alpha}^j\psi_{\beta}^j}{N}\right)^2 - NJ\left(\frac{b^i_{\alpha}\psi^i_{\beta}}{N}\right)\left(\frac{\psi^j_{\alpha}b^j_{\beta}}{N}\right)\left(\frac{\psi^k_{\alpha}\psi^k_{\beta}}{N}\right).
\end{equation}
Here, $\gamma\in \{L,R,L',R'\}$ and $\alpha,\beta$ are summed over the distinct pairs:
\begin{equation}
    (\alpha,\beta)\in\{(L,R),(L,L'),(L,R'),(R,L'),(R,R'),(L',R')\}.
\label{eq:parings}
\end{equation}
We follow the same procedure as in \cref{subsec:saddlept calc Z^2}, where we first introduce the collective fields and integrate out the various $\psi$ and $b$ fields. 
The integral over the $\psi$ and $b$ fields is Gaussian with quadratic form
\begin{equation}
\frac{1}{2}v_i \, M\delta_{ij} v_j ^T \equiv 
\frac{1}{2}v_i \left(\begin{array}{cc}
        A &  B\\
        C & D
    \end{array}\right)\delta_{ij} v_j ^T.
\label{eq:4by4M}
\end{equation}
with
\begin{equation}
v_i =\left( b_L^i,b_R ^i, b_{L^\prime}^i, b_{R^\prime}^i, \psi_L ^i ,\psi_R ^i ,\psi_{L^\prime}^i, \psi_{R^\prime}^i\right)
\end{equation}
and the four times four blocks are given by
\begin{align*}
    &A = \left(
\begin{array}{cccc}
 1 & \frac{\Sigma _{\text{bb}}^{L R}}{2} & \frac{\Sigma _{\text{bb}}^{L L'}}{2} & \frac{\Sigma _{\text{bb}}^{L R'}}{2}  \\
 \frac{\Sigma _{\text{bb}}^{L R}}{2} & 1 & \frac{\Sigma _{\text{bb}}^{R L'}}{2}  & \frac{\Sigma _{\text{bb}}^{R R'}}{2}  \\
 \frac{\Sigma _{\text{bb}}^{L L'}}{2}  & \frac{\Sigma _{\text{bb}}^{R L'}}{2}  & 1 & \frac{\Sigma _{\text{bb}}^{L'R'}}{2}  \\
 \frac{\Sigma _{\text{bb}}^{L R'}}{2}  & \frac{\Sigma_{\text{bb}}^{R R'}}{2}  & \frac{\Sigma _{\text{bb}}^{L' R'}}{2}  & 1 \\
\end{array}
\right),
&B = \left(
\begin{array}{cccc}
 0 & \frac{\Sigma _{\text{b$\psi $}}^{L R}}{2} & \frac{\Sigma _{\text{b$\psi $}}^{L L'}}{2}  & \frac{\Sigma _{\text{b$\psi $}}^{L R'}}{2}  \\
 -\frac{\Sigma _{\text{$\psi $b}}^{L R}}{2}  & 0 & \frac{\Sigma _{\text{b$\psi $}}^{R L'}}{2}  & \frac{\Sigma _{\text{b$\psi $}}^{R R'}}{2}  \\
 -\frac{\Sigma _{\text{$\psi $b}}^{L L'}}{2}  & -\frac{\Sigma _{\text{$\psi $b}}^{R L'}}{2}  & 0 & \frac{\Sigma _{\text{b$\psi $}}^{L' R'}}{2}  \\
 -\frac{\Sigma _{\text{$\psi $b}}^{L R'}}{2}  & -\frac{\Sigma _{\text{$\psi $b}}^{R R'}}{2}  & -\frac{\Sigma _{\text{$\psi $b}}^{L' R'}}{2}  & 0 \\
\end{array}
\right),\\
&C = \left(
\begin{array}{cccc}
 0 & \frac{\Sigma _{\text{$\psi $b}}^{L R}}{2}  & \frac{\Sigma _{\text{$\psi $b}}^{L L'}}{2}  & \frac{\Sigma _{\text{$\psi $b}}^{L R'}}{2}  \\
 -\frac{\Sigma _{\text{b$\psi $}}^{L R}}{2}  & 0 & \frac{\Sigma _{\text{$\psi $b}}^{R L'}}{2}  & \frac{\Sigma _{\text{$\psi $b}}^{R R'}}{2}  \\
 -\frac{\Sigma _{\text{b$\psi $}}^{L L'}}{2}  & -\frac{\Sigma _{\text{b$\psi $}}^{R L'}}{2}  & 0 & \frac{\Sigma _{\text{$\psi $b}}^{L' R'}}{2}  \\
 -\frac{\Sigma _{\text{b$\psi $}}^{L R'}}{2}  & -\frac{\Sigma _{\text{b$\psi $}}^{R R'}}{2}  & -\frac{\Sigma _{\text{b$\psi $}}^{L' R'}}{2}  & 0 \\
\end{array}
\right),
&D = \left(
\begin{array}{cccc}
 0 & \frac{\Sigma _{\psi \psi }^{L R}}{2} & \frac{\Sigma _{\psi \psi }^{L L'}}{2}  & \frac{\Sigma _{\psi \psi }^{L R'}}{2}  \\
 -\frac{\Sigma _{\psi \psi }^{L R}}{2}  & 0 & \frac{\Sigma _{\psi \psi }^{R L'}}{2}  & \frac{\Sigma _{\psi \psi }^{R R'}}{2}  \\
 -\frac{\Sigma _{\psi \psi }^{L L'}}{2}  & -\frac{\Sigma _{\psi \psi }^{RL'}}{2}  & 0 & \frac{\Sigma _{\psi \psi }^{L' R'}}{2}  \\
 -\frac{\Sigma _{\psi \psi }^{L R'}}{2}  & -\frac{\Sigma _{\psi \psi }^{R R'}}{2}  & -\frac{\Sigma _{\psi \psi }^{L' R'}}{2}  & 0 \\
\end{array}
\right).
\end{align*}
The integration over the collective fields is performed in appendix \ref{Appendix:Z4}, with the result (\ref{eq:z4result}). 

On the gravity side the non vanishing of $\langle Z^4\rangle$ can be attributed to wormhole configurations now connecting four boundaries. 
As for $\langle Z^2\rangle$ the same result can be obtained from a saddle point approximation in the large $N$ limit. After a contour deformation these saddle points correspond to pairs of wormholes connecting two boundaries. The saddle points in \cref{sec: Identifying Wormholes} are now ``doubled'' as follows (the $\psi\psi$ label is suppressed on the rhs)
\begin{equation}
\sigma_\psi, g_\psi \rightarrow 
\sigma_{\alpha\beta}, g_{\alpha\beta}
 \label{eq:pairs}
 \end{equation}
 with two $\left( \alpha,\beta\right)$ pairs given by 
 \begin{equation}
 \left( \alpha, \beta\right) \in \left( LR, L^\prime R^\prime\right)\,\,\, \text{or}
 \,\,\, \left( L R^\prime, L^\prime R\right) \,\,\, \text{or} \,\,\,
 \left( L L^\prime, R R^\prime\right) .
 \label{eq:list}
 \end{equation}
 The notation means that the two pairs on the rhs of (\ref{eq:pairs}) take the same values as the $\left( \sigma_\psi, g_\psi\right)$ pair in section \ref{sec: Identifying Wormholes} whereas all other $\sigma_{\alpha\beta}$, $g_{\alpha\beta}$ are zero (the or's in (\ref{eq:list}) are exclusive). The factor three in (\ref{eq:z4result}) can be attributed to the observation that there are three instead of one saddle point and the contribution of each is just the square of its contribution to $\langle Z^2\rangle$ (including one-loop determinants).
 
\section{Half-Wormholes}\label{N1 Phi2}\label{sec: Fixed calculations}
\subsection{Contributions to Non Averaged \texorpdfstring{$\mathbf{Z^2}$}{TEXT}}
Without averaging the expression for $Z^2$ is
\begin{equation}
  Z^2 = \int d^{2N}\psi d^{2N} b\,\,\,
  \text{exp}\left[ \frac{\left(b^i_L\right)^2}{2} +
    \frac{\left(b^i_R\right)^2}{2}
   -\text{i}\sum_{i<j<k} C_{ijk} \left( b^i _L \psi^j _L \psi^k _L
     +b^i _R \psi^j _R \psi^k _R\right)\right] .
\label{eq:noav}
\end{equation}
The authors of \cite{saad2021wormholes} developed an indirect method of approximating this expression by a sum over saddle points. In addition to the previously discussed wormholes there are also contributions due to so called half-wormholes. These correspond to saddle points contributing to $\sqrt{\langle Z^4\rangle}$ but not to $\langle Z^2 \rangle$.
Since (\ref{eq:noav}) clearly factorises into left times right sector the inclusion of half-wormholes restores factorisation.

Again we introduce collective fields by inserting an identity. For the non averaged expression we chose ($d^4G =
dG_{\psi\psi}dG_{bb}dG_{b\psi}dG_{\psi b}$)
\begin{align}
  1 = &\int d^4 G\, \delta\left( G_{\psi\psi} -
    \frac{\psi_L^j\psi_R^j}{N}\right) \delta\left( G_{bb} -
    \frac{b^j_L b^j _R}{N}\right) \delta\left( G_{b\psi} - \frac{b^j
      _L \psi^j _R}{N}\right) \delta\left( G_{\psi b} - \frac{\psi^j
      _L b^j _R}{N}\right)\nonumber \\ 
   & \times \text{exp}\left[-\frac{NJ}{2} \left(
     \frac{b^i_Lb^i_R}{N}\right)
     \left(
           \frac{\psi_L^j \psi_R ^j}{N}\right)^2 +\frac{NJ}{2} G_{bb}
         G_{\psi\psi}^2 \right]\nonumber \\ & \times
         \text{exp}\left[ NJ \left(\frac{b^i _L \psi^i _R}{N}\right) \left(
           \frac{\psi^j_Lb^j_R}{N}\right) \left( \frac{\psi^k_L\psi^k
             _R}{N}\right) -  NJ G_{b\psi}G_{\psi b} G_{\psi\psi}\right]
     \end{align}
The result can then be written as an integral over $\Sigma_{\psi\psi}$ such that,
\begin{equation}\label{eq: z^2 in terms of phi}
    Z^2 = \int d\Sigma_{\psi\psi}\Phi (\Sigma_{\psi\psi}),
\end{equation}
with
\begin{equation}
\Phi\left(\Sigma\right) = \int d^{2N} \psi d^{2N}\, b\, d^4 G \, d \Sigma_{bb} \, d\Sigma_{b\psi}\,  d\Sigma_{\psi b} \, \text{e}^{S} ,
\label{eq:Phi}
\end{equation}
where
\begin{align}
    S &= \frac{1}{2}(b_{L}^i)^2 + \frac{1}{2}(b_{R}^i)^2 + \frac{1}{2}b^i_L\Sigma_{bb}b^i_R + \frac{1}{2}b^i_L\Sigma_{b\psi}\psi^i_R 
    - \frac{1}{2}\psi^i_L\Sigma_{\psi b}b^i_R 
    \nonumber \\ &
    + \frac{NJ}{2}G_{bb}
    (G^{\psi\psi})^2 
    - 2NJ G_{b\psi}G_{\psi b}G_{\psi\psi} 
    - \frac{N}{2}\left(\Sigma_{bb}G_{bb} 
    + \Sigma_{b\psi}G_{b\psi} + \Sigma_{\psi b}G_{\psi b}\right)
    \nonumber \\ &
    -\frac{N}{2}\Sigma G_{\psi\psi}+
    \frac{1}{2}\psi^i_L\Sigma\psi^i_R -\text{i}\sum_{i<j<k} C_{ijk} \left( b^i _L \psi^j _L \psi^k _L
     +b^i _R \psi^j _R \psi^k _R\right)
     \nonumber \\ &
     -\frac{NJ}{2} \left(
     \frac{b^i_Lb^i_R}{N}\right)
     \left(
           \frac{\psi_L^j \psi_R ^j}{N}\right)^2 
           + NJ \left(\frac{b^i _L \psi^i _R}{N}\right) \left(
           \frac{\psi^j_Lb^j_R}{N}\right) \left( \frac{\psi^k_L\psi^k
             _R}{N}\right).
\label{eq:phiac}
\end{align}
In order to check whether a given value of $\Sigma$ belongs to the self-averaging region we compare $\left<\Phi\right>^2$ and $\left<\Phi^2\right>$. 
The average $\langle \Phi\left( \Sigma\right)\rangle$ is the same as (\ref{eq:Phi}) 
with the last three terms in (\ref{eq:phiac}) cancelled.
After contour deformation it is given by (\ref{eq: reduced Z^2 two integrals after grassmann integration}) with the $\sigma_\psi$ integrations dropped and $\sigma_\psi = \text{e}^{\frac{\text{i}\pi}{4}}\Sigma$. For later reference, we notice that it vanishes at $\Sigma = 0$. 
To obtain the expression for $\left<\Phi^2\left(\Sigma\right)\right>$ we begin by first squaring \eqref{eq:Phi} and then average over the coupling which results in,
\begin{equation}\label{eq: Phi^2 eq}
    \left<\Phi^2\left(\Sigma\right)\right> = \int d^{24} G \; d^{22}\Sigma \; d^{2N}\psi\;d^{2N} b\;\text{e}^{I}
\end{equation}
with,
\begin{align}
    I &= \frac{1}{2}\sum_{\alpha}(b_{\alpha}^i)^2 + \frac{1}{2}\sum_{\alpha\beta}\left( b^i_\alpha\Sigma^{\alpha\beta}_{bb}b^i_\beta + \frac{1}{2}b^i_\alpha\Sigma^{\alpha\beta}_{b\psi}\psi^i_\beta - \frac{1}{2}\psi^i_\alpha\Sigma^{\alpha\beta}_{\psi b}b^i_\beta\right) + \frac{NJ}{2}\sum_{\alpha\beta}G^{\alpha\beta}_{bb}(G^{\alpha\beta}_{\psi\psi})^2\nonumber \\
    &
      - 2NJ\sum_{\alpha\beta}G^{\alpha\beta}_{b\psi}G^{\alpha\beta}_{\psi b}G^{\alpha\beta}_{\psi\psi} - \frac{N}{2}\sum_{\alpha\beta}\left(\Sigma^{\alpha\beta}_{bb}G^{\alpha\beta}_{bb}  + \Sigma^{\alpha\beta}_{b\psi}G^{\alpha\beta}_{b\psi} + \Sigma^{\alpha\beta}_{\psi b}G^{\alpha\beta}_{\psi b}\right)\\
    &
     -\frac{N}{2}\Sigma\left( G^{LR}_{\psi\psi}+ G^{L'R'}_{\psi\psi}\right)+
    \frac{1}{2}\Sigma\left( \psi^i_L\psi^i_R + \psi^i_{L^\prime}\psi^i_{R^\prime}\right)
    -\frac{N}{2}{\sum_{\alpha\beta}}^\prime \Sigma^{\alpha\beta}_{\psi\psi} G^{\alpha\beta}_{\psi\psi}+
    \frac{1}{2}{\sum_{\alpha\beta}}^\prime\Sigma^{\alpha\beta}_{\psi\psi} \psi^i_\alpha\psi^i_\beta 
    .\nonumber
\end{align}
Here, $\alpha \in \left\{ L,R,L^\prime, R^\prime\right\}$. In the sum without a prime $\left(\alpha,\beta\right)$ takes values in the list (\ref{eq:parings}) whereas for the primed sums the pairings $LR$ and $L^\prime R^\prime$ are excluded. 
After integrating over the $b$'s and the $\psi$'s this expression looks similar to $\langle Z^4\rangle$ in (\ref{eq:detDexp}) with the difference that $\Sigma_{\psi\psi}^{LR} = \Sigma_{\psi\psi}^{L^\prime R^\prime} =\Sigma$ here and $\Sigma$ is not integrated over. As long as we are interested in $\Sigma$ values which (after contour deformation) happen to be at saddle points of $\langle Z^4\rangle$ the average $\langle \Phi\left(\Sigma\right)^2\rangle$ will be equal to the corresponding saddle point contribution to $\langle Z^4\rangle$. In particular, to $\langle \Phi\left( 0 \right)^2\rangle$ saddle point with non trivial $g_{LL^\prime}, g_{RR^\prime}$ or $g_{LR^\prime}, g_{L^\prime R}$ will contribute (see the discussion at the end of the previous section). Therefore, we find
\begin{equation}
\langle \Phi\left( 0 \right)^2\rangle = 2 \langle Z^2\rangle^2
\end{equation}
 If $\Sigma$ takes non trivial values corresponding to saddle points of $\langle Z^4\rangle$ one has the same contributions to $\langle \Phi\rangle ^2$ and $\langle \Phi^2 \rangle$ at those values.

\subsection{Including Half-Wormholes Restores Factorisation}

Since $\langle \Phi\left( 0\right)\rangle = 0$ the saddle point at $\sigma = 0$ clearly does not contribute to $\langle Z^2 \rangle$. Without averaging everything factorises and all saddle points contributing to $Z^4$ should matter for $Z^2$ as well.  Hence, at $\sigma = 0$  there is a contribution of $\sqrt{\langle\Phi\left( 0 \right)^2\rangle}$ to $Z^2$. The corresponding saddle point is called half-wormhole.
By the same arguments as given in \cite{saad2021wormholes,Peng:2022pfa} one concludes that including half-wormhole contributions suffices to compute the non averaged $Z^2$ and thus restores factorisation. For self-containedness we repeat those arguments following closely \cite{Peng:2022pfa}.

The statement is that the non averaged expression for $Z^2$ equals the averaged expression (dominated by wormhole saddles) plus a half-wormhole contribution

Thus, the fixed coupling expression looks like,
\begin{equation}\label{semi rel}
Z^2 \approx \langle Z^2\rangle + \Phi\left( 0 \right) ,
\end{equation}
in the large $N$ limit. 
In order to check whether \eqref{semi rel} captures essential contributions to the non averaged $Z^2$ we consider the error
\begin{equation}
    \textrm{Error} = Z^2 -\langle Z^2\rangle- \Phi\left( 0 \right) =0.
\end{equation}
The mean error
\begin{equation}
    \left<\text{Error}\right> =  - \left<\Phi\left( 0 \right)\right> 
    =0.
\end{equation}
The mean squared error is given by
\begin{equation}\label{eq: Error2}
    \langle\text{Error.Error}\rangle = \langle\left( Z^2 - \langle Z^2\rangle -\Phi\left( 0 \right)\right)^2\rangle = \langle Z^4\rangle - \langle Z^2\rangle ^2 
    + \langle \Phi\left( 0\right)^2\rangle - 2 \langle Z^2 \Phi\left( 0 
    \right)\rangle .
\end{equation}
Each of these terms is dominated by contributions due to wormhole pairs connecting four boundaries (e.g.\ labelled by $L,R,L^\prime,R^\prime$). However, the number of contributing pairs differs. For instance, for $\langle Z^2\rangle ^2$ only one pairing with $\sigma^{LR}$ and $\sigma^{L^\prime R^\prime}$ non vanishing contributes, for $\langle Z^4\rangle$ there are three such pairings (\ref{eq:list}) whereas for the last two terms only two remain (since $\sigma^{LR}=\sigma^{L^\prime R^\prime}=0$). This leads to
\begin{equation}
    \langle\text{Error.Error}\rangle = \left( 3 -1 + 2 - 4\right) \langle Z^2\rangle ^2
    = 0 .
 \end{equation}
 
\section{Wormholes and Half-Wormholes Break Supersymmetry}\label{SUSY_breaking}

From the susy transformation (\ref{eq:elesusy}) one can deduce how the composite fields transform. Transformations of $G$'s follow directly and the behaviour of $\Sigma$'s can be inferred by imposing invariance on the effective Lagrangians. One finds
\begin{align}
&\delta G^{\alpha\beta}_{\psi\psi} = \epsilon_\alpha G_{b\psi}^{\alpha\beta} + \epsilon_\beta G_{\psi b}^{\alpha\beta} ,\,\,\, \delta G_{b\psi}^{\alpha\beta} = \epsilon_\beta G_{bb}^{\alpha\beta},\,\,\, \delta G_{\psi b}^{\alpha\beta} =\epsilon_\alpha G_{bb}^{\alpha\beta},\,\,\, 
\nonumber \\ &
\delta\Sigma_{b\psi}^{\alpha\beta} = -\epsilon_\alpha \Sigma_{\psi\psi}^{\alpha\beta},
\,\,\, 
\delta\Sigma_{\psi b}^{\alpha\beta} = -\epsilon_\beta \Sigma_{\psi\psi}^{\alpha\beta},
\,\,\,
\delta \Sigma_{bb}^{\alpha\beta} = -\epsilon_\alpha \Sigma_{\psi b}^{\alpha\beta} - \epsilon_\beta \Sigma_{b\psi}^{\alpha\beta}
\end{align}
and the remaining degrees of freedom  are invariant. Here, 
\begin{equation}
\alpha,\beta \in \left\{ L,R\right\} ,\,\,\, \text{respectively}\,\,\, 
\alpha,\beta \in \left\{ L,R,L^\prime, R^\prime\right\} 
\end{equation}
for (\ref{eq:effective}) respectively (\ref{eq:eff4}) and $\alpha \not=\beta$. For the wormhole configuration in section \ref{sec: Identifying Wormholes} we have $G_{\psi\psi}^{LR}\not= 0$.
Therefore $\delta G_{b\psi}^{LR} = \delta G_{\psi b}^{LR} =0$ implies $\epsilon_L = \epsilon_R =0$. For $\langle Z^4\rangle$ the same argument leads to broken supersymmetry. All possible values for $\alpha$ or $\beta$  occur in any of the three options listed in (\ref{eq:list}). Therefore all transformations of $G_{\psi b}^{\alpha \beta}$ and $G_{b \psi}^{\alpha \beta}$ vanish only if all $\epsilon_\alpha =0$. 

\section{Conclusion}\label{sec: Conclusion}

In the presented paper we extended the work of
\cite{saad2021wormholes} to the ${\cal N}=1$ supersymmetric SYK
model. The model was simplified by considering only time independent
fields. Then all moments of the partition function can, in principle,
be computed exactly. At large $N$, the results match a saddle point
approximation if 1-loop determinants are included. We constructed
wormhole and half-wormhole configurations. Wormholes are non trivial
saddle points which reproduce the large $N$ result after contour
deformation. Half-wormholes are saddle points contributing to the
fourth moment but not to the second moment squared. To identify
half-wormholes we expressed the second moment as an integral over one
field (namely $\sigma_{\psi\psi}$). By comparing the expectation value
of the squared integrand to the square of its expectation value one can deduce for
which values of $\sigma_{\psi\psi}$ a given configuration contributes
to the average of $Z^2$. In difference to \cite{saad2021wormholes}  we
were not able to find an expression for any value of
$\sigma_{\psi\psi}$ and depict a self averaging region in the complex
$\sigma_{\psi\psi}$ plane. For our purpose it was sufficient to
consider saddle point values of $\sigma_{\psi\psi}$ contributing to
$\langle Z^4\rangle$. If for a given $\sigma_{\psi\psi}$ there is a contribution to
the second moment of $Z$  there is a wormhole at the corresponding saddle
point and otherwise a half-wormhole. It was argued that the inclusion
of half-wormholes suffices to reproduce $Z^2$ at fixed coupling and
therefore restores factorisation. Wormholes as well as half-wormholes
were shown to break supersymmetry completely.

It would be interesting to extend the results to the full time
dependent supersymmetric SYK model. (For the non supersymmetric SYK
there is a discussion in \cite{saad2021wormholes}.) Another open problem is
to relate half-wormholes to their bulk duals (see
\cite{Garcia-Garcia:2021squ,Goto:2021mbt} for the non
  supersymmetric case). Perhaps, our results could be also confirmed
  by a more direct calculation analogous to
  \cite{Mukhametzhanov:2021nea}. Our observation that wormholes and
  half-wormhole break ${\cal N}=1$ supersymmetry completely could be
  compared to other models with global symmetries such as the complex
  SYK model \cite{Gu:2019jub, Davison:2016ngz} and the ${\cal N} =
  2$ supersymmetric SYK model \cite{SUSY_SYK}.

 \bigskip
\subsection*{Acknowledgments.}
We acknowledge support by
the Bonn Cologne Graduate School of Physics and Astronomy (BCGS). 
\appendix
\section{Computation of \texorpdfstring{$\mathbf{\left<Z^2\right>}$}{TEXT}  \label{sec:int} }
We compute the integral given in (\ref{eq:effective}). First we perform the integrations over the anti-commuting variables. Only terms which are linear in each of them contribute. There are two such terms, the first coming from expanding the first and fourth term in (\ref{eq:effective}) and the second from the last and the next to last terms,
\begin{equation}\label{eq: Z^2 after integrating non commmuting fields}
\left<Z^2\right>_C = I_1 + I_2 ,
\end{equation}
with
\begin{align}
I_1 & = -\frac{N^2J}{32\pi^2}\int \! d^2G d^2 \Sigma\left( 1 -\frac{\Sigma_{bb}^2}{4}\right)^{-\frac{N}{2} -1} 
\!\!\!\!\!\!
\Sigma_{bb}G_{\psi\psi}\left(\frac{\Sigma_{\psi\psi}}{2}\right)^{N-1} \!\!\!\text{e}^{\frac{N}{2}
\left(  J G_{bb} G_{\psi\psi}^2 -\Sigma_{\psi\psi} G_{\psi\psi} - \Sigma_{bb}G_{bb}\right)}
\label{eq:int1}\\
I_2 & = -\frac{N^2}{16\pi^2}\int \! 
d^2G d^2 \Sigma \left( 1 - \frac{\Sigma_{bb}^2}{4}\right)^{-\frac{N}{2}} \left( \frac{\Sigma_{\psi\psi}}{2}\right)^N \text{e}^{\frac{N}{2}
\left(  J G_{bb} G_{\psi\psi}^2 -\Sigma_{\psi\psi} G_{\psi\psi} - \Sigma_{bb}G_{bb}\right)} ,
\label{eq:int2}
\end{align}
where the remaining integrations over commuting quantities are along the real and imaginary axis respectively.
First, the $\Sigma$ integrals are carried out using the general relation
\begin{equation}
\int_{\text{i}{\mathbb R}} d\Sigma \Sigma^k \text{e}^{-\frac{N\Sigma G}{2}} 
= 2\pi \text{i} \left( \frac{2}{N}\right)^{k+1} \left( -1 \right)^k \frac{d^k \delta\left( G\right)}{dG^k} .
\label{eq:gendelta}
\end{equation}
Focusing first on $I_1$, this yields,
\begin{equation}
I_1 = \frac{\left( -1\right)^N J}{N^N}\int\! d^2G\, G_{\psi\psi} \text{e}^{\frac{NJ G_{bb} G_{\psi\psi}^2}{2}} 
\left(  1 - \frac{\partial^2}{N^2 \partial G_{bb}^2} \right)^{-\frac{N}{2} -1}
\frac{\partial^N \delta\left( G_{bb}\right)\delta\left( G_{\psi\psi}\right)}{\partial G_{bb}
\partial G_{\psi\psi}^{N-1}} .
\end{equation}
Performing the $G_{\psi\psi}$ integral yields
\begin{align}
I_1 &= - \frac{1}{N^{N+1}}\int dG_{bb} \frac{1}{G_{bb}} \left. 
\frac{\partial^N \text{e}^{\frac{NJ G_{bb} G_{\psi\psi}^2}{2}}}{\partial G_{\psi\psi} ^N}
\right|_{G_{\psi\psi} = 0}\left( 1 -\frac{\partial^2}{N^2 \partial G_{bb}^2}\right)^{-\frac{N}{2} -1} 
\frac{\partial \delta \left( G_{bb}\right)}{\partial G_{bb}} \nonumber \\
& = \left\{ 
\begin{array}{ c l}
-\frac{N! J^{\frac{N}{2}}}{N^{\frac{N}{2}+1}2^{\frac{N}{2}}\left(\frac{N}{2}\right)!}
\int dG_{bb} G^{\frac{N}{2}-1}_{bb} \left( 1 -\frac{\partial^2}{N^2 \partial G_{bb}^2}\right)^{-\frac{N}{2} -1} 
\frac{\partial \delta \left( G_{bb}\right)}{\partial G_{bb}} & \text{for\ even\ $N$}\\
0 & \text{for\ odd\ $N$} \end{array}\right.
\end{align}
For even $N$ we expand the integral in powers of derivatives. Only the term for which 
the number of derivatives equals $\frac{N}{2}-1$ contributes. 
From the expansion of the integrand we can get only odd powers of derivatives implying 
that for a non vanishing result $N$ has to be a multiple of four. We obtain
\begin{equation}
I_1 =\left\{ \begin{array}{cl}\left( -1\right)^{\frac{N}{4}-1}\frac{ N!\left(\frac{N}{2}-1\right)! J^{\frac{N}{2}}}{N^{N-1} 2^{\frac{N}{2}}\left( \frac{N}{2}\right)!}
\left(\begin{array}{c} -\frac{N}{2}-1 \\ \frac{N}{4}-1 \end{array} \right) & \text{for\ } 4 \mid N ,\\
0 & \text{for\ } 4 \nmid N .\end{array}\right.
 \end{equation}
Here, the last factor denotes the binomial coefficient which we compute by analytic continuation
\begin{equation}
\left(\begin{array}{c} -\frac{N}{2}-1 \\ \frac{N}{4}-1 \end{array} \right) =
\frac{\Gamma\left( -\frac{N}{2}\right) }{\Gamma\left( \frac{N}{4}\right) \Gamma\left( -\frac{3N}{4} +1 \right)}
= \left( -1\right)^{1-\frac{N}{4}}\frac{\left( \frac{3N}{4}-1\right) !}{\left( \frac{N}{4}-1\right)! \left(\frac{N}{2}\right)!}.
\end{equation}
In the last step we have deformed negative integer arguments of Gamma functions by adding $\epsilon$. Then we have replaced the corresponding Gamma functions by means of 
$$\Gamma\left( x\right) \Gamma\left( 1- x\right) = \frac{\pi}{\text{sin} \pi x}$$ 
and taken $\epsilon$ to zero in the end. This leads to
\begin{equation}
\left( \begin{array}{c} -M \\ k \end{array} \right)
= \frac{\left( -1\right)^k \left( M + k-1\right)! }{k! \left( M-1\right)!} ,
\label{e:binomial}
\end{equation}
with $M$ taken to be positive and integer. 
Finally, we use Stirling's approximation for large $N$ and obtain
\begin{equation}
I_1 \approx\left\{ \begin{array}{cl}  2 \sqrt{\frac{2}{3}}\,2^{-N} 3^{\frac{3N}{4}} J^\frac{N}{2} \text{e}^{-N}
& \text{for\ } 4 \mid N , \\
0 & \text{for\ } 4 \nmid N .\end{array}\right.
 \label{eq:1stt}
 \end{equation}
The computation of the second term is quite similar. After integrating out $\Sigma_{\psi\psi}$ and $\Sigma_{bb}$ we obtain
\begin{equation}
I_2 = N^{-N}  \int d^2 G \,\text{e}^{\frac{NJ G_{bb} G_{\psi\psi}^2}{2}}\left( 1 - \frac{\partial^2}{N^2\partial G_{bb}^2}\right)^{-\frac{N}{2}} \frac{\partial^N}{\partial G_{\psi\psi}^N} \delta\left( G_{bb}\right) \delta\left( G_{\psi\psi}\right) .
\end{equation}
Performing the $G_{\psi\psi}$ integral results in
\begin{equation}
I_2 = \left\{ \begin{array}{cl} \frac{N^{-\frac{N}{2}} N! J^{\frac{N}{2}}}{2^{\frac{N}{2}}\left(\frac{N}{2}\right)! }
\int dG_{bb}\, G_{bb}^{\frac{N}{2}}\left( 1 - \frac{\partial^2}{N^2\partial G_{bb}^2}\right)^{-\frac{N}{2}}
\delta\left( G_{bb}\right) & \text{for even\ } N ,\\
0 & \text{for odd\ } N .
\end{array}\right. 
\end{equation}
Since the integrand gives rise only to even numbers of derivatives non vanishing results can be only obtained if the power $N/2$ is even, i.e.\ $N$ is a multiple of four. Proceeding in the same way as for $I_1$ we find that in the 
large $N$ limit 
\begin{equation}
I_2 = I_1 ,
\end{equation}
with $I_1$ given in (\ref{eq:1stt}).
\section{One-Loop Determinants \label{ap:loop}}
Expanding the Lagrangians in (\ref{eq: reduced lagrangians 1 and 2}) to second power in fluctuations (denoted e.g.\ by $\delta \sigma_\psi$) around the saddle point \eqref{eq: L1 saddles} and \eqref{eq: L2 saddles} with $m=0$
\begin{equation}
{\cal L}_{1/2} = {\cal L}_{0,1/2} - \frac{1}{2}c S_{1/2}  c^T  ,
\end{equation}
where ${\cal L}_{0,1/2}$ denotes the saddle point value and
\begin{equation}
c = \left( \delta g_\psi, \delta\sigma_\psi\right) \,\,\, ,
\end{equation}
For ${\cal L}_1$ we have,
\begin{equation}
S = \left(\begin{array}{cc}
\frac{5\text{i}J\sqrt{3}N}{2} &\frac{\text{i}N}{2}\\
\frac{\text{i}N}{2} &-\frac{\text{i}N}{2J\sqrt{3}}
\end{array}\right) 
\end{equation}
integrating over the fluctuations provides a factor 
\begin{equation}
2 \pi \sqrt{\frac{1}{\det S}} = \frac{2\pi}{N}\sqrt{\frac{2}{3}}.
\end{equation}
For ${\cal L}_2$ we get the same value for the one-loop determinant contribution.
\section{Computing \texorpdfstring{$\mathbf{\langle Z^4\rangle}$}{TEXT}}\label{Appendix:Z4}

After replacing the elementary $\psi^i$ and $b^i$ by collective fields the mean of $Z^4$ is
\begin{align}
\left< Z^4\right> =\left( \frac{1}{2\pi}\right)^{12} \int d^{24} \Sigma d^{24} G \left(s\text{det} M\right)^{-N/2} &
\nonumber \\&\hspace*{-2.4in}\times
\text{e}^{-\frac{N}{2}\sum_{\alpha , \beta}\left( \Sigma_{\psi\psi}^{\alpha\beta}
G_{\psi\psi}^{\alpha\beta} + \Sigma_{bb}^{\alpha\beta}G_{bb}^{\alpha\beta} +
\Sigma_{b\psi}^{\alpha\beta}G_{b\psi}^{\alpha\beta} + \Sigma_{\psi b}^{\alpha\beta}G_{\psi b}^{\alpha\beta}
 + J G_{bb}^{\alpha\beta}\left(G_{\psi\psi}^{\alpha\beta}\right)^2 -2J G_{b\psi}^{\alpha\beta} G_{\psi b}^{\alpha\beta} G_{\psi\psi}^{\alpha\beta} \right)}
\label{eq:eff4}
\end{align}
where $M$ is defined in (\ref{eq:4by4M}) and the sum includes all 
configurations listed in (\ref{eq:parings}). Performing all $G^{\alpha\beta}_{b\psi}$ integrals yields
\begin{align}
\left< Z^4\right>_C =\left( \frac{1}{2\pi}\right)^{12} \int d^{24} \Sigma d^{24} G \left(s\text{det} M\right)^{-N/2}\prod_{\alpha,\beta}\left( \frac{N}{2} \Sigma_{b\psi}^{\alpha\beta}- NJ G_{\psi b}^{\alpha\beta}G_{\psi\psi}^{\alpha\beta}\right) &
\nonumber \\&\hspace*{-3in}\times
\text{e}^{-\frac{N}{2}\sum_{\alpha , \beta}\left( \Sigma_{\psi\psi}^{\alpha\beta}
G_{\psi\psi}^{\alpha\beta} + \Sigma_{bb}^{\alpha\beta}G_{bb}^{\alpha\beta} +
  \Sigma_{\psi b}^{\alpha\beta}G_{\psi b}^{\alpha\beta}
+ J G_{bb}^{\alpha\beta}\left(G_{\psi\psi}^{\alpha\beta}\right)^2 \right)}
\label{eq:z4grasout}
\end{align}
The product in the first line gives rise to $2^6$ terms,
\begin{equation}
\left< Z^4\right>_C =\sum_{i=1}^{64} K_i .
\label{eq:terms}
\end{equation}
We focus in the following on the first term containing a product of all six $\Sigma_{b\psi}^{\alpha\beta}$ and call the result $K_1$.  For this term one can easily perform the $G_{\psi b}^{\alpha\beta}$ integrals, all Grassmann odd quantities in $M$ can be set to zero. We arrive at
\begin{equation}
K_1 = \frac{1}{\left( 2 \pi\right)^{12}} \left( \frac{N}{2}\right)^{12} \int d^{12} \Sigma d^{12} G \left( \frac{\det D}{\det A}\right)^{\frac{N}2}
\text{e}^{-\frac{N}{2}\sum_{\alpha , \beta}\left( \Sigma_{\psi\psi}^{\alpha\beta}
G_{\psi\psi}^{\alpha\beta} + \Sigma_{bb}^{\alpha\beta}G_{bb}^{\alpha\beta} 
 + J G_{bb}^{\alpha\beta}\left(G_{\psi\psi}^{\alpha\beta}\right)^2  \right)} .
 \end{equation}
To simplify the calculation we deform integration contours partially by integrating $\Sigma_{bb}^{\alpha\beta}$ along the real axis and $G_{bb}^{\alpha\beta}$ along the imaginary axis. Then $G_{bb}^{\alpha\beta}$ integrations yield delta functions
$$ \left( 2 \pi \text{i} \right)^6\left(\frac{2}{N}\right)^6 \delta^{(6)}\left( \Sigma_{bb}^{\alpha\beta} + J \left(G_{\psi\psi}^{\alpha\beta}\right)^2\right) $$
which we exploit when performing the $\Sigma_{bb}^{\alpha\beta}$ integrals. The remaining variables carry all $\psi\psi$-subscripts which we drop in the following (and move $\alpha\beta$ down),
\begin{equation}
K_1 = -\frac{1}{\left( 2 \pi\right)^6} \left( \frac{N}{2}\right)^6\int d^6 \Sigma d^6 G \left( \frac{\det D}{\det \tilde{A}}\right)^{\frac{N}{2}} \text{Exp}\left[ -\frac{N}{2}\sum_{\alpha\beta} \Sigma_{\alpha\beta} G_{\alpha\beta}\right].
\end{equation}
Here,
\begin{eqnarray}
 \det\tilde{A} &= &\det A \big\vert_{\Sigma_{bb}^{\alpha\beta} \to - 
J \left(G_{\alpha\beta}\right)^2}\nonumber \\ 
& = &1 -\frac{J^2}{4}\sum_{\alpha\beta} G_{\alpha\beta}^4 -
\frac{J^3}{4}\left( G_{LR}^2G_{LL^\prime}^2 G_{RL^\prime}^2 
+ G_{LR}^2 G_{LR^\prime}^2 G_{RR^\prime}^2 
+ G_{LL^\prime}^2 G_{LR^\prime}^2 G_{L^\prime R^\prime}^2 \right. \nonumber\\
& & \left. +G_{RL^\prime}^2 G_{RR^\prime}^2 G_{L^\prime R^\prime}^2 \right)+ \frac{J^4}{16}\left( G_{LR}^4 G_{L^\prime R^\prime}^4 + G_{LL^\prime}^4 G_{RR^\prime}^4 
+ G_{LR^\prime}^4 G_{RL^\prime}^4 \right.
\\
& & \left.
- 2 G_{LR}^2 G_{LL^\prime}^2 G_{RR^\prime}^2 G_{L^\prime R^\prime}^2 
- 2 G_{LR}^2 G_{LR^\prime}^2 G_{RL^\prime}^2 G_{L^\prime R^\prime}^2
-2 G_{L L^\prime}^2 G_{L R^\prime}^2 G_{R L^\prime}^2 G_{R^\prime R^\prime}^2 
\right)
\nonumber
\end{eqnarray}
Performing the $\Sigma_{\alpha\beta}$ integrals using (\ref{eq:gendelta}) yields
\begin{equation}
K_1 = \int d^6 G \delta^{(6)}\left( G_{\alpha\beta}\right) 
\det \tilde{D}^{\frac{N}{2}} \det \tilde{A}^{-\frac{N}{2}}
\label{eq:K1}
\end{equation}
with
\begin{equation}
\tilde{D} = D \big\vert_{\Sigma^{\alpha\beta} \to -\frac{2}{N}\frac{\partial}{\partial G_{\alpha\beta}}} .
\end{equation}
Notice, that e.g 
$$\det \tilde{D}\det\tilde{A} \not= \det \tilde{D}\tilde{A }$$
since the components of the two matrices do not commute. All derivatives should be kept to the left. The appearing powers of the determinants can be expanded as follows
\begin{align}
\det \tilde{D}^{\frac{N}{2}} & = \frac{1}{N^{2N}}\left( \frac{\partial^2}{\partial G_{LR} \partial G_{L^\prime R^\prime}} 
-\frac{\partial^2}{\partial G_{LL^\prime} \partial G_{R R^\prime}}
+ \frac{\partial^2}{\partial G_{LR^\prime} \partial G_{R L^\prime}}\right)^N
\nonumber \\ & 
= \frac{1}{N^{2N}}\sum_{l_1 + l_2 + l_3 =N}\frac{N!}{l_1 ! l_2 ! l_3 !} \left( -1\right)^{l_2} \frac{\partial^{2N}}{\partial G_{LR}^{l_1} 
\partial G_{L^\prime R^\prime}^{l_1}\partial G_{L L^\prime}^{l_2}
\partial G_{R R^\prime}^{l_2} 
\partial G_{L R^\prime}^{l_3}\partial G_{R L^\prime}^{l_3}},
\label{eq:detDexp}
\end{align}
whereas
\begin{equation}
\det \tilde{A}^{-\frac{N}{2}} = \sum_{k=0}^\infty 
\left( \begin{array}{c} -\frac{N}{2} \\ k \end{array}\right)
\sum_{\sum_{i=1}^{16} k_i = k}\frac{k!}{\prod_{i=1}^{16} k_i !}\left( -1\right)^x \, 2^y \, J^z \prod_{\alpha\beta} G_{\alpha\beta}^{2k_{\alpha\beta}} ,
\end{equation}
with
\begin{equation*}
x = \left(\sum_{i=1}^{10} +\sum_{i=14}^{16}\right) k_i ,\,\,\, y = \left(-2  \sum_{i=1}^{10}   
-4\sum_{i =11}^{13}  -3 \sum_{i=14}^{16}\right) k_i,\,\,\, z = \left(2\sum_{i=1}^6  + 
3 \sum_{i=7}^{10} + 4 \sum_{i=11}^{16}\right) k_i ,
\end{equation*}
\begin{align}
k_{LR} = 2 k_1 + k_7 +k_8 + 2k_{11}+ k_{14} + k_{15},  &\quad 
k_{LL^\prime} = 2 k_2 + k_7 + k_9 +2 k_{12} + k_{14} + k_{16}, \nonumber \\
k_{LR^\prime} = 2 k_3 + k_8 + k_9 +2 k_{13} + k_{15} + k_{16}, & 
\quad k_{RL^\prime} = 2 k_4 + k_7 + k_{10} +2 k_{13} + k_{15} + k_{16},
\nonumber \\
k_{RR^\prime} = 2k_5 + k_8 +k_{10} +2 k_{12} + k_{14} +k_{16},& \quad
k_{L^\prime R^\prime} = 2 k_6 + k_9 + k_{10} + 2 k_{11} + k_{14} + k_{15} .
\nonumber
\end{align}
Notice that for the $k^{\text{th}}$ term the total degree of the appearing monomials 
$2\sum k_{\alpha\beta}\geq 4k$. All terms in $\det \tilde{D}^{\frac{N}{2}}$ are differential operators of order $2N$. Therefore, from the infinite sum over $k$ at most terms with $k\leq N/2$ contribute to the integral (\ref{eq:K1}). One obtains
\begin{align}
K_1 = &\frac{1}{N^{2N}}\sum_{k=0}^{N/2} \left( \begin{array}{c} -\frac{N}{2} \\ k \end{array}\right) \sum_{\sum k_i = k}\left(-1\right)^x 2^y J^z N!\, \left(2k_{LR}\right)!\,\left( 2 k_{L R^\prime}\right)!\,\left( 2 k_{L L^\prime}\right)!\, \frac{k!}{\prod k_i !}
\nonumber \\
& \times\delta_{k_{LR}k_{L^\prime R^\prime}}
\delta_{k_{L R^\prime}k_{RL^\prime}}\delta_{k_{LL^\prime}k_{R R^\prime}}
\delta_{k_{LR} + k_{L R^\prime} + k_{L L^\prime }-N/2} .
\label{eq:z4closed}
\end{align}
To make further progress we focus on particular contributions. First we will be interested in the contribution with $k_{LR}=k_{L^\prime R^\prime} = N/2$ and the rest of the $k_{\alpha\beta} =0$. The Kronecker symbols are non vanishing for (we consider again $N$ to be a multiple of four)
\begin{equation}
k_1 = k_6 = l\,\,\, ,\,\,\, k_{11} = \frac{N}{4} -l\,\,\, \text{with\ } l \in \left\{ 0, \ldots , \frac{N}{4}\right\}
\end{equation}
and all other $k_i$'s are zero.
The sum over $k$ and the  $k_i$'s becomes 
\begin{equation}
\sum_{l=0}^{N/4} \frac{\left( \frac{3N}{4} + l -1 \right)!}{l!^2 \left( \frac{N}{4} - l \right)!} \left( - 1\right)^{l +\frac{N}{4}} = \frac{ \left( \frac{3N}{4} -1\right)!^2}{\left( \frac{N}{4}\right)!^2 \left(\frac{N}{2}-1\right) !}
\label{eq:sumnon}
\end{equation}
Plugging this into (\ref{eq:z4closed}) and taking the large $N$ limit we find
\begin{equation}
\left. {K_{1}}\right|_{k_{LR}= k_{L^\prime R^\prime}=N/2} =
\frac{8}{3}\times 2^{-2N}3^{\frac{3N}{2}} J^N \text{e}^{-2N} .
\end{equation}
This particular result could have been obtained without performing the non trivial sum (\ref{eq:sumnon}) by setting the corresponding $G_{\alpha\beta}=0$ in $\tilde{A}$ and factorising the determinant before the expansion. 

\begin{figure}[h]
\begin{center}
\includegraphics{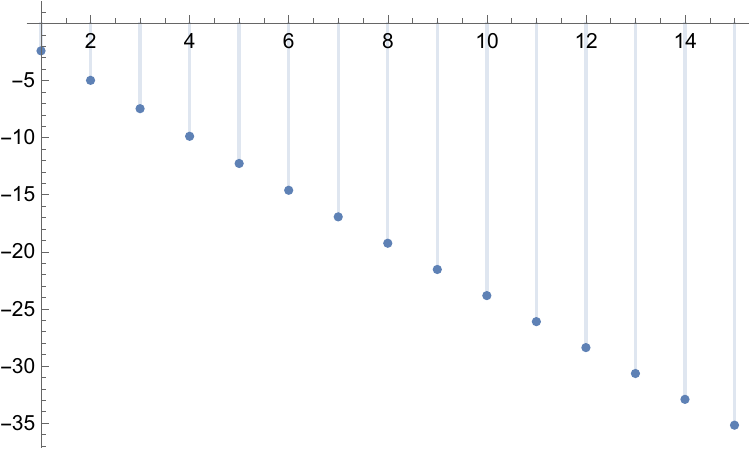}
\end{center}
\caption{$\text{Log} \left( \left. {K_{1}}\right|_{k_{LR}= k_{L^\prime R^\prime}=k_{LR^\prime} =k_{RL^\prime} =\frac{N}{4}}/ \left. {K_{1}}\right|_{k_{LR}= k_{L^\prime R^\prime}=N/2}\right)$ as a function of $N/8$\label{fig:ratio}}
\end{figure}

Next, we consider the contribution with
\begin{equation}
k_{LR} = k_{L^\prime R^\prime} = k_{LR^\prime} =k_{RL^\prime} = \frac{N}{4}
\end{equation}
and all other $k_{\alpha\beta}=0$. Non vanishing contributions can be parameterised as
\begin{align}
& k_1 = k_6 = l ,\,\,\, k_3 = k_4 = p, \,\,\, k_{11} = \frac{N}{8} -l-q,\,\,\, k_{13} =\frac{N}{8} - p-q,\nonumber \\& k_{15}=2q ,\,\,\, l,p,q \in \left\{ 0,\ldots, \frac{N}{8}\right\}
\end{align}
with all other $k_i=0$. We assumed that $N$ is a multiple of $8$ (for odd $4N$ only odd choices of $k_{15}$ contribute).
We were not able to obtain a closed expression for the resulting triple sum.
 Therefore we just plot $\text{Log} \left( \left. {K_{1}}\right|_{k_{LR}= k_{L^\prime R^\prime}=k_{LR^\prime} =k_{RL^\prime} =\frac{N}{4}}/ \left. {K_{1}}\right|_{k_{LR}= k_{L^\prime R^\prime}=N/2}\right)$ in figure \ref{fig:ratio}.
We see that contributions with more than one pair of $k_{\alpha\beta}$ being large are exponentially suppressed. A similar behaviour  was also observed in other models
\cite{saad2021wormholes,Peng:2022pfa}. Assuming that all contributions in (\ref{eq:terms}) give either the same result or are subleading in the large $N$ limit we obtain finally\footnote{Here, a factor of three is included taking into account contributions from equivalent configurations with $k_{RL^\prime}= k_{LR^\prime}= N/2$ respectively $k_{LL^\prime} = k_{RR^\prime}=N/2$.}
\begin{equation}
\left< Z^4\right> \approx \alpha \times  8\times 2^{-2N}3^{\frac{3N}{2}} J^N \text{e}^{-2N},
\end{equation}
where $\alpha$ is the number of contributions from the sum in (\ref{eq:terms}).
Finally, we are going to argue that $\alpha = 4$. 
Also for all other terms in (\ref{eq:terms}) one can perform first the $G_{bb}^{\alpha\beta}$ integrations such that one is left only with $G_{\psi\psi}^{\alpha\beta}$ and $\Sigma_{\psi\psi}^{\alpha\beta}$ integrals. The $\Sigma_{\psi\psi}^{\alpha\beta}$ integrals are performed using (\ref{eq:gendelta}).
Now we assume that all terms contributing at leading order for large $N$
correspond to terms containing just derivatives with respect to two different $G_{\alpha\beta}$ and not more\footnote{In the previous computation we observed that from the power in (\ref{eq:detDexp}) no mixed products contributed at leading order.}.
Let us focus on terms containing derivatives with respect to $G_{LR}$ and 
$G_{L^\prime R^\prime}$. Then all $G_{\psi\psi}^{\alpha\beta}$ integrations for $\left(\alpha ,\beta\right) \notin \left\{ \left( L,R\right),\left( L^\prime , R^\prime\right)\right\}$ are performed by setting $G_{\psi\psi}^{\alpha\beta} =0$. This implies that from the second term in the product in (\ref{eq:z4grasout}) only those with $\left(\alpha ,\beta\right) \in \left\{ \left( L,R\right),\left( L^\prime , R^\prime\right)\right\}$ contribute. In these cases the integral factorises into products of integrals discussed in appendix \ref{sec:int}, where we found two equal contributions. Hence, $\alpha = 2^2 =4$. Finally, we obtain
\begin{equation}
\left< Z^4\right> = 3 \left< Z^2\right>^2 .
\label{eq:z4result}
\end{equation}
As for the non supersymmetric SYK model \cite{saad2021wormholes} this relation is consistent with a Gaussian distribution of $Z$ itself. 

\bibliographystyle{utphys.bst}
\bibliography{bibliography}
\end{document}